\documentclass{article}

\usepackage{PRIMEarxiv}

\usepackage[utf8]{inputenc} 
\usepackage[T1]{fontenc}    
\usepackage{hyperref}       
\usepackage{url}            
\usepackage{booktabs}       
\usepackage{amsfonts}       
\usepackage{nicefrac}       
\usepackage{microtype}      
\usepackage{lipsum}
\usepackage{graphicx}
\usepackage{float} 
\usepackage{amsmath} 
\usepackage{cleveref} 
\graphicspath{{media/}}     
\usepackage{graphicx}
\usepackage{caption}
\usepackage{subcaption}
\usepackage[colorinlistoftodos]{todonotes}

\makeatletter
\def\blfootnote{\xdef\@thefnmark{}\@footnotetext}
\makeatother

\begin{document}
\raggedbottom

\title{The rising costs of training frontier AI models}

\author{\textnormal{Ben Cottier\textsuperscript{1}} 
\and \textnormal{Robi Rahman\textsuperscript{1,2}}
\AND \textnormal{Loredana Fattorini\textsuperscript{2}}
\and \textnormal{Nestor Maslej\textsuperscript{2}} 
\and \textnormal{Tamay Besiroglu\textsuperscript{1}} 
\and \textnormal{David Owen\textsuperscript{1}}}

\date{}
\maketitle

\begin{abstract}
The costs of training frontier AI models have grown dramatically in recent years, but there is limited public data on the magnitude and growth of these expenses. This paper develops a detailed cost model to address this gap, estimating training costs using three approaches that account for hardware, energy, cloud rental, and staff expenses. The analysis reveals that the amortized cost to train the most compute-intensive models has grown precipitously at a rate of $2.4\times$ per year since 2016 (90\% CI: $2.0\times$ to $2.9\times$). For key frontier models, such as GPT-4 and Gemini, the most significant expenses are AI accelerator chips and staff costs, each costing tens of millions of dollars. Other notable costs include server components (15-22\%), cluster-level interconnect (9-13\%), and energy consumption (2-6\%). If the trend of growing development costs continues, the largest training runs will cost more than a billion dollars by 2027, meaning that only the most well-funded organizations will be able to finance frontier AI models.\blfootnote{$^1$Epoch AI. $^2$Stanford University.}
\end{abstract}

\section{Introduction}
\label{sec:sec1}

The large and growing cost of training state-of-the-art AI models has become an important issue in the field of AI~\cite{maslej2023artificial}. Improving AI capabilities demand exponential increases in computing power, as evidenced by both economic analysis~\cite{thompson2022importance} and the discovery of empirical scaling laws, which show that model performance improves with more parameters and training data~\cite{kaplan2020scaling,hoffmann2022training}. Dario Amodei, CEO of the AI lab Anthropic, has stated that frontier AI developers are likely to spend close to a billion dollars on a single training run this year, and up to ten billion-dollar training runs in the next two years~\cite{klein2024podcast}. Given this trend, some innovations, particularly those requiring large-scale training, may become inaccessible to all but the most well-funded organizations.

Although it is widely known that training the largest ML models is expensive, until recently there were few concrete estimates of training costs in the public domain. In collaboration with Epoch AI, the 2024 AI Index presented one of the most comprehensive datasets to date, estimating the costs of training runs based on cloud rental prices~\cite{maslej2024artificial}. We build on that work with a more in-depth account of hardware, energy and R\&D staff costs for both training runs and experiments, as well as a more detailed analysis of how costs are increasing over time. To our knowledge, our study is the most thorough analysis of model development costs to date.

Our methods are built upon a comprehensive database of notable machine learning models~\cite{epoch2023pcdtrends}, and informed by interviews with industry experts. We consider three complementary approaches to measuring the cost of frontier models. The first approach estimates the hardware capital expenses (CapEx) amortized over the final training run, along with the cost of hardware energy consumption. By considering AI accelerator chips, other server hardware, networking hardware, and energy separately, this approach can provide more accurate training costs. We find that the most expensive publicly-announced training runs to date are OpenAI's GPT-4 at \$40M and Google's Gemini Ultra at \$30M. Among frontier models, defined as models within the top 10 most compute-intensive models when they are released, we find that training has become $2.4\times$ more expensive per year since 2016 (90\% CI: $2.0\times$ to $2.9\times$).

\begin{figure}[htb]\centering
  \includegraphics[width=0.8\textwidth]{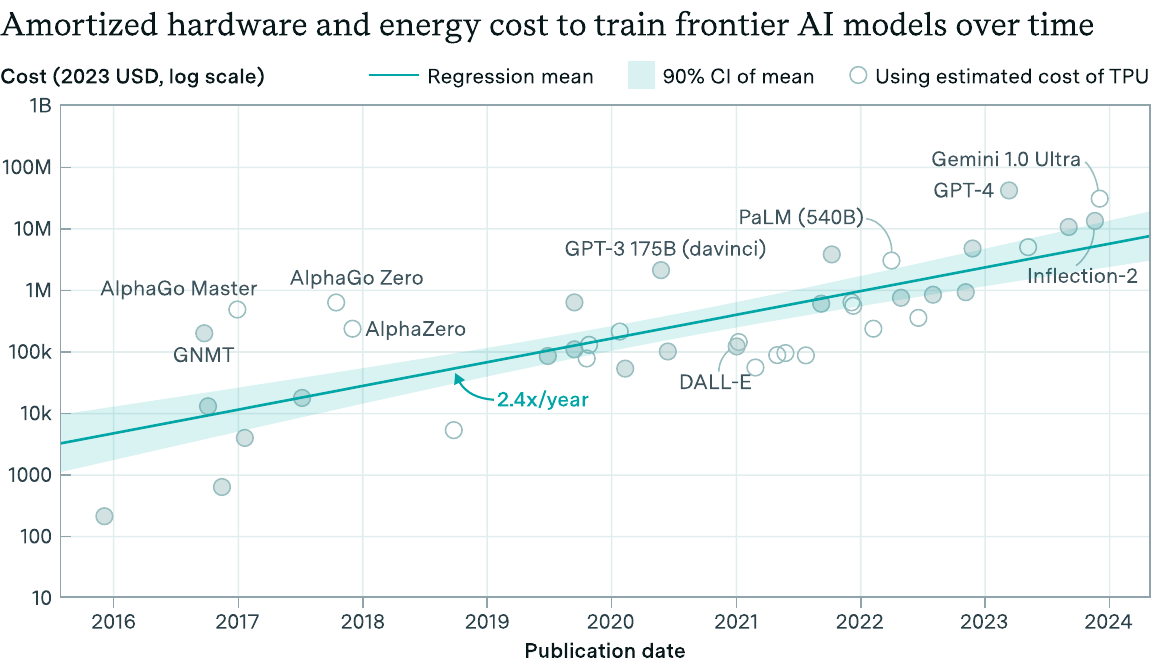}
\caption{Amortized hardware cost plus energy cost for the final training run of frontier models. The selected models are among the top 10 most compute-intensive for their time. Amortized hardware costs are the product of training chip-hours and a depreciated hardware cost, with 23\% overhead added for cluster-level networking. Open circles indicate costs which used an estimated production cost of Google TPU hardware. These costs are generally more uncertain than the others, which used actual price data rather than estimates.}
\label{fig:f1}
\end{figure}

We then compare this approach to the cloud-price approach that was first presented in the AI Index~\cite{maslej2024artificial}. Instead of estimating hourly compute costs in detail, the cloud-price approach simply uses historical rental rates from cloud platforms. The cloud-price approach shows a similar growth rate ($2.5\times$ per year with a 90\% CI of $2.1\times$ to $3.1\times$), but yields costs that are about twice as large on average. We expect the cloud-price approach to overestimate frontier model costs, since model developers usually either own or have private rental agreements for their training hardware. Using both approaches helps validate our estimate of cost growth, while also highlighting the uncertainty of individual costs.

Our third and most in-depth approach breaks down hardware, energy, and R\&D staff costs over the entire development of the model (i.e. both experiments and training). We select four especially notable models for this approach---GPT-3, OPT-175B, GPT-4, and Gemini Ultra. For these models, we find that R\&D staff costs including equity are between 29\% and 49\% of the total amortized cost. Computing hardware makes up 47--64\%, while energy comprises only 2--6\%. However, if we exclude equity the fraction for R\&D staff drops to 19--33\%, and the fractions of computing hardware costs and energy rise to 61--76\% and 2--7\% respectively.

By taking into account hardware purchase costs, energy costs, and the more opaque costs of R\&D labor, our analysis provides a clearer picture of the true costs of AI development. This sheds light on not only current costs but also the economic hurdles that lie ahead as AI continues to scale.

All of our results can be reproduced using the code and data available at \url{https://github.com/epoch-research/training-cost-trends}.

\section{Methodology}
\label{sec:sec2}

\subsection{Datasets and frontier model selection}

Our investigation draws upon the Notable AI Models database, which documents 796 notable models across the history of machine learning~\cite{epoch2023pcdtrends}. Key details captured for each model include training compute, dataset size, and parameter count. To focus on the largest-scale models, we initially filtered the database to models that had training compute estimates and that were published on or after 1 October 2015 (the start of the large-scale ML era according to \cite{sevilla2022compute}) and up to 31 December 2023. This resulted in 276 selected models. For these models, we recorded the training time, hardware type and quantity, and utilization rate sourced from each model's original publication, where possible.

For our main results, we examined 41 models that were historically at the frontier of compute. Specifically, we filtered for models that were in the top 10 of training compute as of their release.\footnote{We excluded models that are fine-tuned versions of a separately listed model, to avoid double-counting costs.} \Cref{sec:ap1} provides further details on this selection procedure and a comparison to three alternative methods. 

In addition to the data on machine learning models, we compiled a dataset of historical hardware prices, allowing us to estimate training costs. This price dataset contained cloud rental prices and hardware purchase prices for 24 different hardware models (e.g. NVIDIA A100) between 2015 and 2023. In total there were 142 entries, 52 of which were purchase prices and 90 of which were cloud rental prices.

\subsection{Amortizing the cost of hardware for training}
\label{sec:amort}

To estimate the cost of hardware for a training run, we first calculated the cost to acquire the necessary accelerator chips (GPUs/TPUs), servers, and networking hardware. This involved looking up historical prices for GPUs, or estimating production costs for TPUs. Further details are provided in \Cref{sec:ap2}. 

Hardware normally remains available for future use after a training run finishes, but its value depreciates over time due to hardware progress. We amortized the cost of a training run based on this depreciation. Specifically, we depreciated the value of hardware at a rate of $r=0.14$ orders of magnitude per year, based on the growth rate of ML GPU price-performance \cite{epoch2023trendsinmachinelearninghardware}.\footnote{This growth rate measures improvement at 32-bit precision. One-time improvements from lower-precision and tensor number formats would make the rate faster, but this was not estimated. This also assumes that hardware improves continuously. In reality, hardware improves in increments with each new release.} To get the value of the hardware at the start of training, we used the following formula:
\begin{gather*}
\text{Start value per chip} =
\dfrac{\text{Acquisition cost per chip}}{\exp\Big(\big[
\text{Training start date} - \text{Hardware availability date}\big]\cdot r\ln{10}\Big)}
\end{gather*}

where the difference in dates is in units of years. For example, if the training run starts one year after hardware is acquired, the start value is approximately 72\% of the acquisition cost. We neglected the impact of hardware failures on depreciation, as the effect seemed small compared to hardware progress. We provide evidence for that in \Cref{sec:ap4}.

After finding the initial value of the hardware, the amortized cost of the training run is then the portion of value that is lost during the training run. However, the training time is often difficult to determine, due to a lack of public information. More often, we were able to estimate the number of chip-hours: the product of the training time and the number of chips. So we substituted chip-hours for the training time and the number of chips, using a linear approximation. This led to our final formula for amortized training cost:
\begin{gather*}
\text{Amortized training cost} \approx \text{Start value per chip}\times\frac{\text{Training chip-hours}}{(365 \times 24) \text{ hours/year}}\times r\ln{10}
\end{gather*}
Up until Section \ref{sec:totaldevcost}, our results only account for the chip-hours of the final training run. In Section \ref{sec:totaldevcost}, we scale up the chip-hours to account for all experiments towards developing an AI model. Although the amortized cost model involves several estimates and approximations, our results are robust to reasonable changes in the method (see \Cref{sec:ap4} for further methodological details and \Cref{sec:ap73} for the sensitivity analysis).

\subsection{Hardware energy consumption cost}

In addition to the capital costs of hardware, we also considered the cost of energy consumed by hardware during model training. We estimated this using the following formula:
\begin{gather*}
\begin{split}
\text{Total energy cost of training} =\ 
& \text{Energy cost rate (\$/kWh)} \times
\text{Hardware TDP (kW)}\ \times
\\
& \text{Average power to TDP ratio (\%)}\times
\text{Data center PUE}\times\text{Training chip-hours (h)}
\end{split}
\end{gather*}
where TDP is thermal design power and PUE is power usage effectiveness, which accounts for the overhead of data center power distribution and cooling. We selected the energy cost rate by year, hardware TDP by the hardware type, average power to TDP ratio by the hardware manufacturer, and the data center PUE by the ML model developer. These were set based on hardware manufacturers' literature. However, some parameters such as average power to TDP ratio could not be found in technical specifications and had to be estimated. For references and method details, see \Cref{sec:ap5}.

\subsection{Cloud compute cost}

While the amortized hardware CapEx + energy approach is a bottom-up method that accounts for hardware and energy costs, cloud rental prices offer a simpler method. Many AI labs rely on cloud computing services to train their models, and the associated costs are often more readily available and easier to estimate. By comparing our bottom-up estimates with those derived from cloud rental prices, we can validate our approach and provide a more comprehensive picture of AI training costs. The cloud approach also allows estimates of model training cost that are not possible using the amortized hardware CapEx + energy approach and our data. However, note that the cloud approach can overestimate the cost of models whose developers used their own hardware rather than renting compute from a cloud provider.

To estimate training costs from cloud rental prices, we used the following formula:
\begin{gather*}
\text{Total cost} = \text{Price per chip-hour} \times
\text{Training chip-hours}
\end{gather*}

The price per chip-hour was obtained from our hardware price database, which includes prices for various hardware types, cloud providers, and rental dates. We matched the hardware type and publication date of each ML model with the most appropriate price, using the developer of the ML model to determine the most likely cloud provider (e.g., Google Cloud for Google labs, Microsoft Azure for OpenAI). See \Cref{sec:ap6} for further details.

\subsection{Total amortized model development cost}

Although the final training run ultimately determines an AI model's capabilities and impact, the research and development surrounding it is crucial. We therefore used a third approach that considers all of the compute that went into model development, as well as the cost of R\&D staff developing the model. Since this approach was more time-intensive, and relied on having a list of contributors to estimate R\&D staff cost, we applied it to just four models: GPT-3, OPT-175B, GPT-4, and Gemini Ultra.

To estimate the compute cost over model development---including experiments, failed attempts, evaluation and fine-tuning---we applied a multiplicative factor to the final training run compute. We estimated this factor based on evidence about the development of GPT-3, OPT-175B and BLOOM, as well as the general AI infrastructure at Meta. \Cref{sec:devcompute} provides further details. Based on this, we sampled the factor from a log-normal distribution with a 90\% CI of 1.2x to 4x, meaning that total compute for model development is 1.2x to 4x larger than the final training run.

\subsubsection{R\&D staff costs}

Research and development (R\&D) staff costs are an often-neglected component of the total cost of developing ML models. These costs include the salaries and equity compensation of the researchers, engineers, and managers in the project, but excludes operations staff and data center employees. We set out to better quantify these costs for a few selected models to see how significant they are relative to the hardware costs.

We estimated total annual compensation of R\&D personnel by multiplying the estimated full-time equivalent workload per contributor by their compensation and by the total time spent on model development. Since these parameters were all quite uncertain, we sampled from log-normal distributions over each parameter.

For full-time equivalent workers, we were informed by the type and number of contributors listed on the research paper. For all models except Gemini Ultra, we sampled full-time equivalent workloads from a 90\% credible interval of 5\% to 80\% FTE for each contributor, resulting in a median of 20\%. For Gemini Ultra, we used different workloads for each type of contributor listed~\cite[pp. 66--69]{geminiteam2024gemini}.

For compensation, we were informed by company-specific data from \url{https://www.levels.fyi/} and \url{https://aipaygrad.es/}. From levels.fyi, we used data for Google Software Engineers from level 3 to level 8. From aipaygrad.es, we used the overall statistics for all companies and all roles (researchers, engineers and managers). After averaging the two sources, base salaries were modeled with a 90\% CI of \$140K to \$160K, and equity with a 90\% CI of \$35K to \$490K. We applied an overhead factor of 1.25x to 1.4x to base salaries to account for taxes and benefits~\cite{weltman2019}, resulting in total compensation with a 90\% CI of \$210K to \$690K and a median of \$330K.

Actual staff compensation may vary significantly between AI labs. The chosen estimate of compensation may be particularly unreliable for small and early companies, such as OpenAI in its earlier years, where there are many uncertainties about how to value equity compensation. However, these numbers serve as a reasonable baseline, and our estimates provide a useful starting point to analyze R\&D labor costs.

\newpage

\section{Results}
\label{sec:sec3}
\subsection{Amortized training costs of frontier models have grown by 2.4x per year since 2016}

\begin{figure}[htb]\centering
  \includegraphics[width=0.8\textwidth]{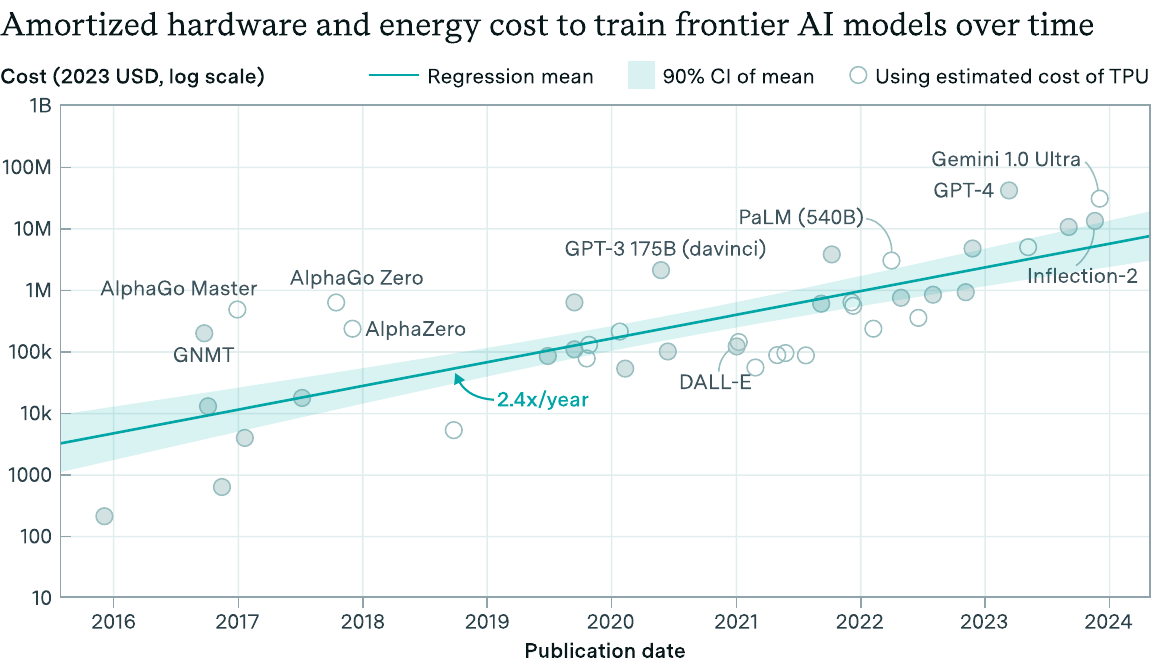}
\caption{(Reproduction of \Cref{fig:f1} for convenience.) Amortized hardware cost plus energy cost for the final training run of frontier models. The selected models are among the top 10 most compute-intensive for their time. Amortized hardware costs are the product of training chip-hours and a depreciated hardware cost, with 23\% overhead added for cluster-level networking. Open circles indicate costs which used an estimated production cost of Google TPU hardware. These costs are generally more uncertain than the others, which used actual price data rather than estimates.}
\label{fig:f1reproduced}
\end{figure}

The amortized training costs of frontier models have increased by a factor of 2.4x per year since 2016. This is the result of the preferred amortized hardware CapEex + energy approach, shown in \Cref{fig:f1reproduced}. \Cref{tab:t1} compares this to the cloud approach, which yields a similar growth rate of $2.5\times$ per year. The growth rate is also similar if we vary hardware depreciation or training start date within reasonable limits (see \Cref{sec:depsensapp}). However, the growth rate rises to 2.9x per year if we exclude TPUs, which have more uncertain costs than publicly-sold GPUs.

\begin{table}[bht]
\centering
\renewcommand{\arraystretch}{1.4}
\resizebox{\textwidth}{!}{%
\begin{tabular}{p{3.75cm}ccccc}
\toprule
\textbf{Approach} & \textbf{$\boldsymbol{N\times}$ increase per year} & \textbf{OOMs/year} & \textbf{Doubling Time (months)} & \textbf{R-squared} & \textbf{N} \\
\midrule
Amortized hardware \newline CapEx + energy & 2.4 [2.0, 2.9] & 0.38 [0.29, 0.47] & 9 [8, 12] & 0.58 & 41 \\
Amortized hardware \newline CapEx + energy---no TPUs & 3.0 [2.4, 3.7] & 0.47 [0.37, 0.57] & 8 [6, 10] & 0.77 & 22 \\
Renting from the cloud \newline & 2.5 [2.1, 3.1] & 0.40 [0.32, 0.48] & 9 [7, 11] & 0.66 & 36 \\
\bottomrule
\end{tabular}
}
\vspace{8pt}
\caption{Cost growth rates based on log-linear regression, for different cost estimation approaches. All approaches select the top 10 most compute intensive models at the time of model release. $N$ refers to the number of relevant observations. Based on a two-sided $t$-test adjusted for correlation of residuals, the growth rates for amortized hardware capex + energy and cloud are not significantly different $(p=0.13)$. However, when the costs of models trained with estimated TPU production costs are excluded, the growth rate rises significantly to 2.9x per year ($p<0.01$). OOMs/year: orders of magnitude per year. Square brackets: 90\% confidence interval.}
\label{tab:t1}
\end{table}

\begin{figure*}[t]\centering
    \centering
    \includegraphics[width=0.8\textwidth]{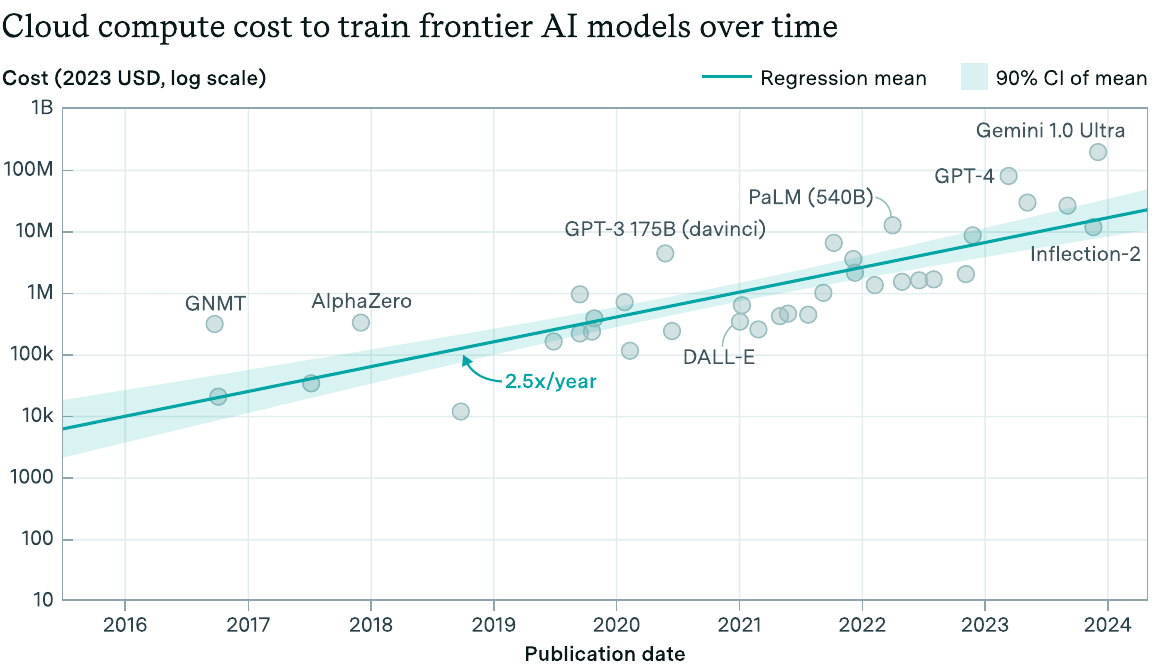}
    \caption{Estimated cloud compute costs for the final training run of frontier models. The selected models are among the top 10 most compute-intensive for their time. The costs are the product of the number of training chip-hours and a historical cloud rental price.}
    \label{fig:f2}
\end{figure*}

Estimating costs from cloud rental prices, although less representative of actual costs, has the advantage of simplicity. The cloud cost approach also helps to check the robustness of the amortized hardware CapEx + energy approach. \Cref{fig:f2} shows the trend of cloud compute cost to train models among the top 10 most compute-intensive as of their release. Note that some of these estimates previously appeared in the 2024 AI Index report~\cite{maslej2024artificial}. We find that the cost of training models based on cloud rental prices has grown by $2.5\times$ per year since 2016, with a 90\% CI of $2.1\times$ to $3.1\times$. This is consistent with the amortized hardware CapEx + energy approach, as shown in \Cref{tab:t1}. This shows that our trend estimates are robust to two different ways of estimating prices per chip-hour.

Overall, these results suggest that the cloud approach is valid for estimating \textit{growth rates} in compute costs, and has the advantage of simplicity. However, public cloud rental prices are less reliable for \textit{individual} model costs when the model developer owns the hardware or has a special partnership with a cloud provider.

\subsection{The trend suggests that the most expensive publicly announced model will cost one billion dollars to train by the start of 2027}

The growth rate in training cost indicates how rapidly AI investment is scaling. We can use this growth rate to extrapolate the cost of the largest training run. Currently, GPT-4 has the largest amortized hardware and energy cost, at \$40M. GPT-4 was published in March of 2023 \cite{openai2024gpt4}. This implies that, at a growth rate of $2.4\times$ per year, the most expensive publicly announced model by the start of 2027 will cost about \$1 billion.

Whether this cost is justified hinges on how profitable the resulting AI model is---but parts of the AI industry believe it is worthwhile. The CEO of the AI lab Anthropic has claimed that close to a billion dollars will already be spent on a single training run in 2024 (implying an amortized cost), which is even sooner than the historical trend suggests \cite{klein2024podcast}.

\subsection{Hardware acquisition costs are one to two orders of magnitude higher than amortized costs}

It's important to distinguish the amortized cost of the hardware used for training, which is spread over the useful lifetime of the hardware, and the acquisition cost of purchasing that hardware outright. The choice of which cost to consider depends on the purpose of the analysis. Amortized costs are more relevant for understanding the economics of training and deploying models over an extended period, while acquisition costs give a sense of the capital barriers to entry and financial risks involved in developing such models.

\begin{figure}[htb]\centering
\includegraphics[width=0.8\textwidth]{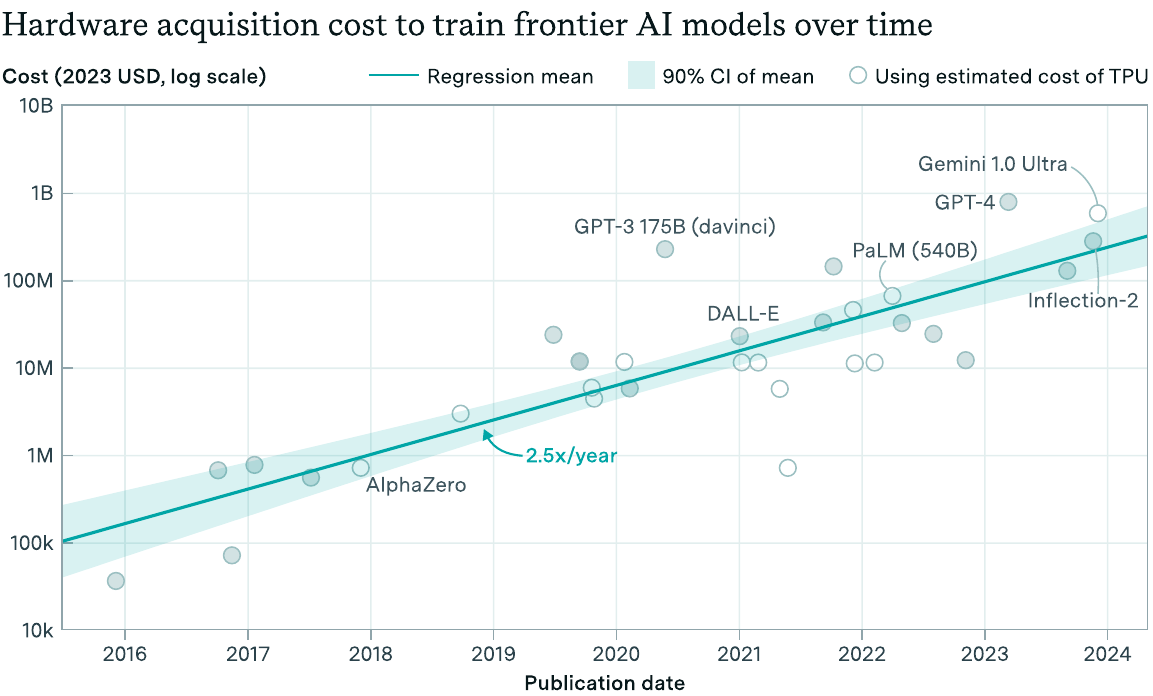}
\caption{Estimated hardware acquisition costs to train frontier models. The selected models are among the top 10 most compute-intensive for their time. The costs are the product of the number of servers and the earliest available server price, with about 23\% overhead added for cluster-level networking hardware.}
\label{fig:f3}
\end{figure}

To illustrate the difference between amortized hardware cost and acquisition cost, \Cref{fig:f3} shows the acquisition costs we were able to estimate using hardware purchase prices and training hardware quantities. Since this is the \textit{up-front} cost of acquiring the hardware, the costs are one to two orders of magnitude larger than amortized hardware costs. For example, we estimate that it cost \$800M to acquire the hardware used to train GPT-4, compared to \$40M for the amortized hardware CapEx + energy cost. The ratio between the two depends on when and for how long the model is trained.

Based on 40 estimates of acquisition cost, we find a growth rate of $2.5\times$ per year (90\% CI: $2.1\times$, $3.0\times$). This is slightly faster than the rate of $2\times$ per year suggested by amortized costs ($2.4\times$ per year) divided by training times ($1.2\times$ per year)~\cite{epoch2024time}. The discrepancy is due to different AI models appearing in each analysis, highlighting a source of sensitivity in our results.

\subsection{Half of amortized hardware CapEx + energy cost is for AI accelerator chips}

\begin{figure}[htbp]
    \centering
    \includegraphics[width=0.85\textwidth]{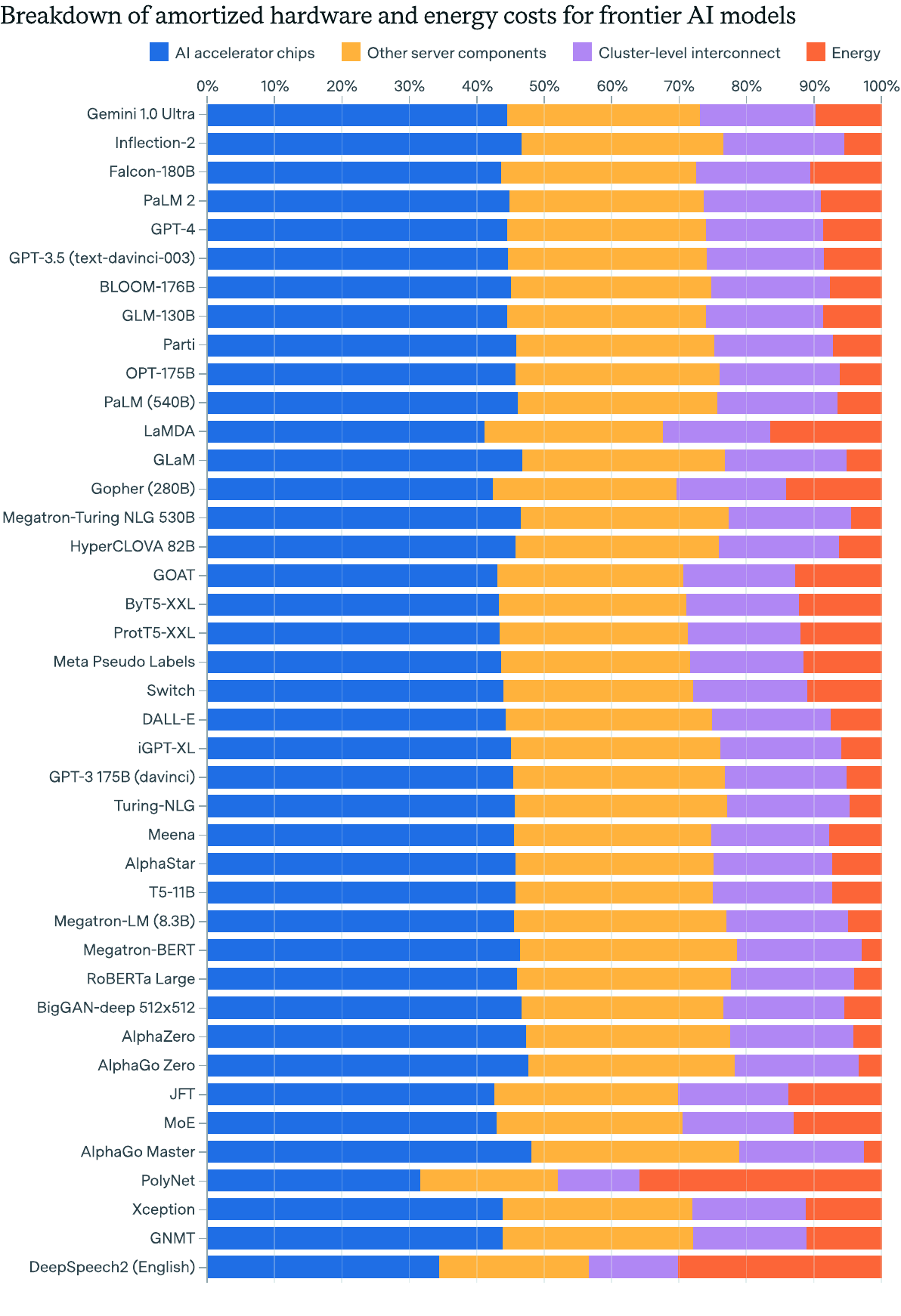}
    \caption{The percentage of the amortized hardware CapEx + energy estimates made up by different hardware and energy costs. Note that the breakdown across models is approximate. Cluster-level interconnect is assumed to be a constant 19\% fraction of the cluster CapEx, and the proportion of server components is based on only three comparisons between NVIDIA DGX\ server prices and single GPU prices (see \Cref{sec:ap2} for details). The energy costs are more specific, varying with the number of training chip-hours and the hardware (see \Cref{sec:ap5}).}
    \label{fig:f4}
\end{figure}

Breaking down the components of amortized hardware CapEx + energy in \Cref{fig:f4}, we find that on average, 44\% goes toward AI accelerator chips. The rest of the server (including markup) makes up 29\% of the cost, while cluster level interconnect makes up 17\%.

Energy makes up the remainder of costs, averaging 9\% but varying across models. Although this is a small fraction, it corresponds to rapid growth in energy use and power requirements over time. The trend in power requirements is provided in \Cref{sec:powerapp}.

Note that this breakdown does not include all costs associated with an AI supercomputer. Other costs include the data center infrastructure besides servers and networking, as well as data center personnel and maintenance.

\subsection{R\&D staff are a significant fraction of costs over the whole model development process}
\label{sec:totaldevcost}

\begin{figure}[htb]
    \centering
    \begin{subfigure}[t]{1\textwidth}
        \centering
	\includegraphics[width=0.8\textwidth]{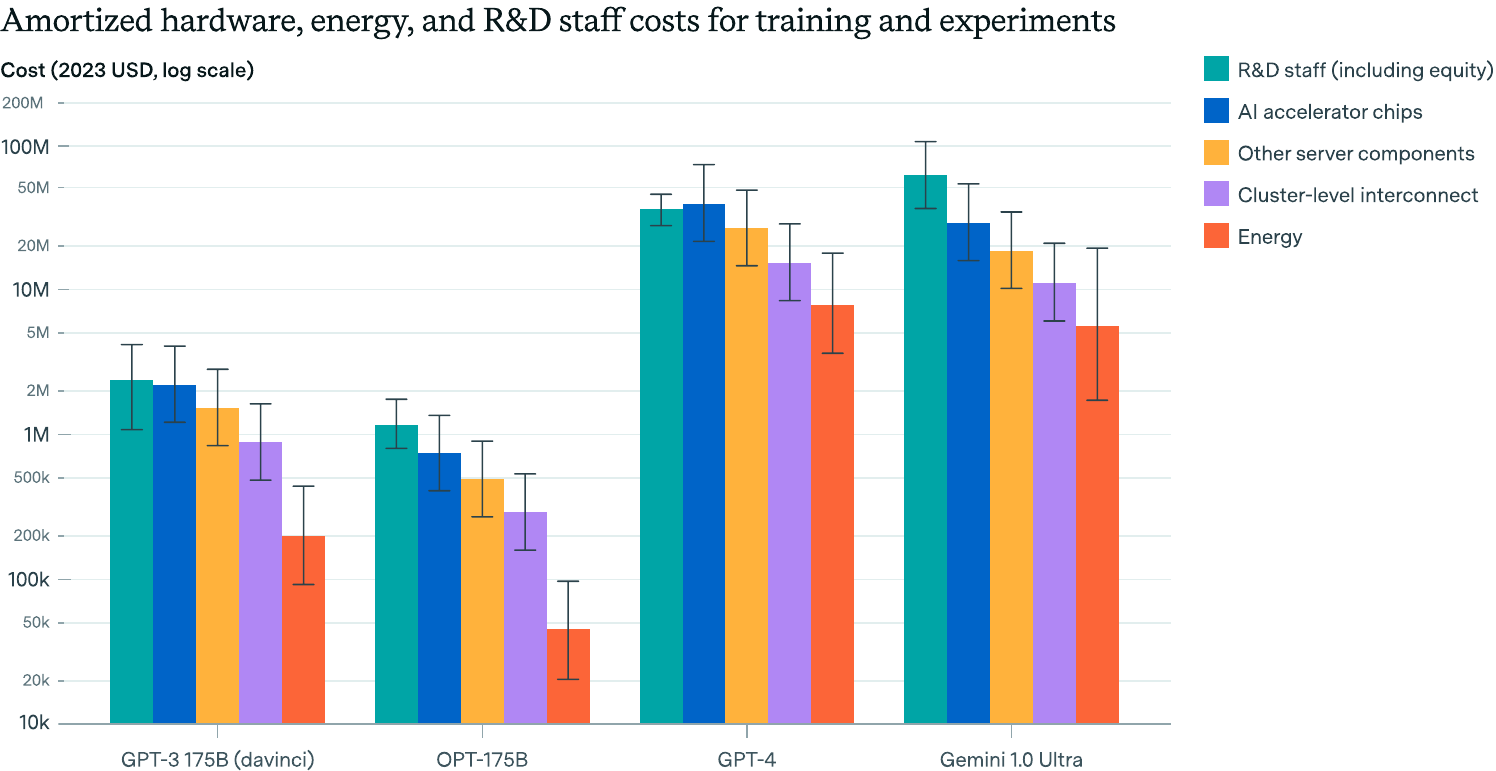}
        \caption{}\label{fig:f6a}
    \end{subfigure}\medskip%
    
    \begin{subfigure}[t]{1\textwidth}
        \centering
	\includegraphics[width=0.7\textwidth]{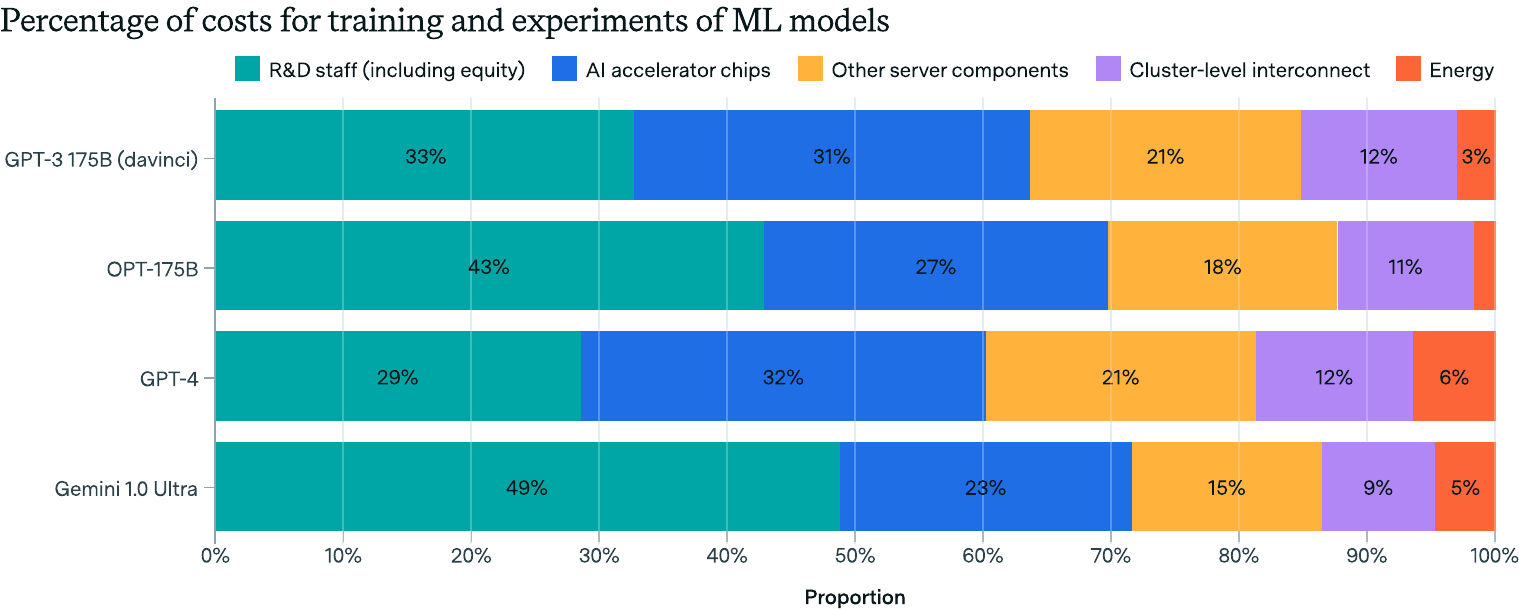}
        \caption{}
        \label{fig:f6b}
    \end{subfigure}
\caption{(a) Breakdown of total amortized model development costs for selected models. Hardware costs are amortized to the total number of chip-hours spent on experiments and training, while R\&D staff costs cover the duration of development from initial experiments to publication. Error bars indicate 90\% credible intervals, while the main bar values are medians. (b) Costs components as a percentage of the total, based on median estimates.}
\label{fig:f6}
\end{figure}

We now use our third cost estimation approach to examine how the cost of labor from researchers and engineers compares to the amortized cost of compute. Unlike the previous approaches, which only measured the cost of the final training run, this approach counts compute usage throughout model development including experiments, fine-tuning and evaluation. \Cref{fig:f6} shows the cost breakdown for GPT-3, OPT-175B (notable as a GPT-3 replication attempt by a team at Meta AI), the original GPT-4 model by OpenAI, and the original Gemini 1.0 Ultra model by Google DeepMind.

We find that when equity is included, R\&D staff costs make up between 29\% and 49\% of total amortized model development costs, depending on the model. Excluding equity, the fraction decreases to 21\% to 33\% (see  \Cref{sec:ap74} for additional plots). Notably, this fraction does not change much from GPT-3 to GPT-4, which spans three and a half years of AI progress. The number of reported contributors increased from 25 for GPT-3~\cite{brown2020language} to 284 for GPT-4~\cite{openai2024gpt4}, while the amortized hardware cost over the whole model development increased from \$4M to \$90M. However, due to the limited data, we caution against extrapolating the fraction of R\&D staff costs to future frontier models.

Gemini Ultra has the highest fraction of R\&D staff cost at 49\%, but we expect this is unusually high among frontier models. Firstly, Gemini Ultra was trained on Google TPUs, which are cheaper for Google than buying other accelerators, and this makes the hardware cost relatively low. Secondly, our methodology is limited by deriving the number of full-time equivalent staff from the reported number of contributors, for which Gemini had 941---much higher than GPT-4 at 284 contributors. Though we assumed a very small contribution from the 428 people under the ``Contributors'' role---a median full-time equivalent of about 1\%---the estimate may still be too high.

On the compute side, we find that amortized hardware cost makes up 47--64\% of the full model development cost, while energy comprises only 2--6\%. With equity excluded from R\&D costs, the fraction of hardware cost and energy cost rise to 61--76\% and 2--7\% respectively. Note that while energy consumption is a small fraction of total cost, this doesn't entail that power requirements are not a challenge in frontier AI development. Regulatory and logistical hurdles to secure power supplies may cause bottlenecks in the coming years, but we leave that topic to future work.

\section{Discussion}

\subsection{Implications}
The rapid growth in AI training costs will have a major impact on the future of AI development. Our findings suggest that if the current trend of 2.4x per year growth continues, then the amortized cost of frontier training runs will exceed one billion dollars by 2027. Given the potential bias in our estimates' absolute values, this may happen even sooner---as suggested by cloud-price costs, and news reporting on training costs \cite{klein2024podcast}. If realized, this level of investment is likely to drive rapid advances in AI capabilities, given the track record of scaling up AI models.

However, only a handful of large companies and government institutions have the financial resources to operate at this frontier. This concentration of AI development could limit the range of perspectives and approaches considered, especially from academia and broader society. Both AI developers and policymakers must grapple with the rapid AI advances brought on by increasing investment, as well as the tradeoffs involved in the concentration of AI development. On one hand, having few key players at the frontier could make it easier for them to coordinate on responsible AI development. On the other hand, this raises concerns about a lack of public oversight for such a powerful technology.

\subsection{How to estimate training costs}
We used two approaches to estimate the cost of final training runs: the amortized hardware CapEx + energy approach, and the cloud rental price approach. These two approaches produced consistent estimates of the growth rate in training cost over time. However, the approaches diverged on individual costs: the cloud costs were twice as large on average. We recommend using the amortized hardware CapEx + energy approach for frontier models wherever it's feasible, because it accounts for the lower costs in practice for large training runs, and can be broken down into components.

Our third approach adds the cost of R\&D staff, as well as the compute cost of experiments, evaluations, and fine-tuning involved in model development. To our knowledge, we present the first detailed estimates of these costs for GPT-3, OPT-175B and Gemini Ultra. Moreover, our results suggest that R\&D staff costs were a major component of total costs for these frontier models. Although this is the most comprehensive of the three approaches, further data collection and evidence on the AI development process are needed before we can recommend it as the default.

\subsection{Limitations}
While our study provides valuable insights into the growth of AI training costs, there are important limitations. The analysis relies on publicly available information, which may lead to biases or gaps in the dataset. Cost estimation methods are subject to uncertainties due to factors such as hardware depreciation rates and pricing dynamics. Moreover, our methods neglect several costs that are potentially significant, including the data center infrastructure apart from the training cluster, and the acquisition of data for model training.

Our results may also have limited generality. The trends observed for the selected frontier models may not generalize to the broader AI landscape, or specific AI domains such as language modeling. Rapid  innovation could also lead to large gains in hardware and software efficiency that are difficult to predict from historical data. Further research on all of these unknowns would help refine our insights, and inform evidence-based strategies to respond to growing financial barriers in ML.

\section{Conclusion}

In this paper we used three approaches to analyze the cost of training ML models at the frontier. The first two approaches---one based on hardware purchase prices and energy costs, the other based on cloud rental prices---indicate that the amortized cost of compute for these training runs has grown by around 2.4x per year (90\% CI: 2.0x to 2.9x) since 2016. This shows the large role of investment in driving AI progress.

Breaking down the total amortized model development cost for selected frontier models (GPT-3, OPT-175B, GPT-4 and Gemini Ultra), we found that R\&D staff are a major component, making up 29--49\% of the total. This motivates further research on the scaling of R\&D labor with computing power.

The rapid exponential growth of costs over eight years suggests that growth is unlikely to stall in the next few years. However, frontier AI labs appear to face non-trivial challenges to scaling further. One such challenge is securing enough power capacity for increasingly large computing clusters. Analyzing potential bottlenecks such as this is an important topic for future work.

The rapid increase in AI investment is likely to drive major advances in AI capabilities. Given that total model development costs at the frontier are already over \$100 million, these advances may only be accessible to the largest companies and government institutions. The concentration of such a powerful technology among a few key players raises questions about responsible development and deployment. Both AI developers and policymakers must engage with these issues and consider the tradeoffs involved. The stakes are high---decisions made now about the governance and trajectory of AI could have profound consequences for society.

\section*{Acknowledgements}

We thank Bartosz Podkanowicz for assisting with data collection, Luke Frymire for assisting with ground truth cost verification, and Josh You for copyediting. We thank Igor Molybog, Yafah Edelman and Horace He for helpful conversations about the training process and requirements for creating large ML models. We thank Konstantin Pilz, Yafah Edelman, Tom Davidson, Isabel Juniewicz, Carl Shulman, Jaime Sevilla, Aleksandar Kostovic, Tim Fist, Haydn Belfield, Alan Chan, David Patterson, Mauricio Baker, Erich Grunewald, and Cullen O’Keefe for feedback on drafts.

This study was supported by a grant from the AI Index based out of the Stanford Institute for Human-Centered Artificial Intelligence. The cloud compute cost estimates shown here previously appeared in the 2024 Stanford AI Index Report.

\bibliography{references.bib}

\clearpage
\appendix

\section{Training cost estimation}
\label{sec:training_cost}
\subsection{Hardware acquisition cost}
\label{sec:ap2}

The frontier AI models in our dataset were trained on clusters of many GPU or TPU chips (or ``chips'' for short). We set out to estimate the total cost of the chips, servers, and networking equipment in these clusters, which we call the hardware acquisition cost. This cost is calculated as follows:
\begin{gather*}
\text{Hardware acquisition cost} = \text{Acquisition cost per chip} \times \text{Number of chips}
\end{gather*}
Where ``Acquisition cost per chip'' accounts for the GPU or TPU chip itself, other server costs (CPUs, memory, chip-to-chip networking, and markup), and the cost of server-to-server networking equipment. \Cref{tab:ta6} shows how we calculated this quantity depending on what was known about the training hardware.

\begin{table}[H]
\centering
\renewcommand{\arraystretch}{1.4}
\begin{tabular}{ll}
\toprule
\textbf{Known information} & \textbf{Formula for ``Acquisition cost per chip''} \\
\midrule
Single GPU price & $\text{GPU chip price} \times \text{Chip-to-server factor} \times \text{Server-to-cluster factor}$ \\
GPU server price & $\text{GPU server price} \:/\: \text{GPUs per server} \times \text{Server-to-cluster factor}$ \\
Chips are TPUs & $\text{TPU chip cost} \times \text{Chip-to-server factor} \times \text{Server-to-cluster factor}$ \\
\bottomrule
\end{tabular}%
\vspace{8pt}
\caption{The formula to estimate ``Acquisition cost per chip'', depending on what is known about the training hardware.}
\label{tab:ta6}
\end{table}

If the training hardware was a GPU, we looked up the earliest known price linked to that GPU. If the price we found was for a single GPU, we multiplied that price by a ``chip-to-server'' cost factor (detailed later). If the price we found was instead for a DGX server, we divided that price by the number of GPUs per server.\footnote{HGX servers are more suited to large-scale, customized AI supercomputers, but we found very little information on their pricing, so we used DGX pricing.} Finally, if the training hardware was a Google TPU, we used the ``Geometric mean'' production cost estimate from \Cref{tab:ta2}. This represents the cost of the TPU chip itself, which we then multiplied by a chip-to-server cost factor.

We calculated chip-to-server cost factors based on known DGX and single-GPU prices near release, using the formula (DGX cost)/(8$\times$GPU cost). We were able to estimate this for the NVIDIA P100 $(1.54\times)$, V100 $(1.69\times)$, and A100 $(1.66\times)$. For other NVIDIA chips, and for TPUs, we used the mean of these three known factors $(1.64\times)$. We assumed that the DGX server prices included the cost of chip-to-chip interconnect switches and transceivers.\footnote{Some of the NVIDIA product pages where we found hardware prices listed NVSwitches as components, but it was unclear whether NVLink cables for chip-to-chip links were included.} We did not account for financing, i.e. the interest paid on a loan to purchase the hardware up-front.

Once we had the cost per chip for a single server, we added the cost of server-to-server networking equipment. We used an estimate by Kostovic (forthcoming), based on the reference architecture of the NVIDIA H100 SuperPOD~\cite{nvidia2023superpodrefarch}. According to this estimate, approximately 19\% of the total cost of the SuperPOD goes towards cluster-level interconnect for configurations with less than 4096 GPUs, and 20\% for 4096 GPUs and above, due to an additional third layer of switches. Consistent with these figures, another expert in AI hardware estimated a range of 10\% to 20\% for A100-based clusters with an Infiniband network~\cite{fist2024}.

For simplicity, we assumed that 19\% of the hardware acquisition cost was for server-to-server networking equipment. We therefore multiplied the cost per chip for a single server by a ``server-to-cluster'' factor of $100\%/(100\%-19\%)\approx1.23\times$, resulting in the final ``Acquisition cost per chip''. We assumed that the overhead factor is accurate for TPU servers as well as GPU servers, though we have substantial uncertainty about this. In reality, the proportion of costs varies with the cluster architecture and size.

\subsection{Cost of Google TPUs}
\label{sec:ap3}

Tensor Processing Units are a class of proprietary AI accelerator hardware developed by Google, and used in their internal computing projects and employed in Google Cloud datacenters~\cite{jouppi2017datacenter}. These chips are not available for sale, but some of them can be rented on the cloud. Since they have never been sold, there are no available purchase prices, which makes it more difficult to estimate the amortized capital expenses for Google Brain, DeepMind, and other Google labs.

To estimate the cost of TPUs used by Google labs, we aggregated two approaches. The first approach estimates TPU manufacturing costs based on a bill of materials (BOM) for the NVIDIA H100 GPU. We consider this a low-end estimate, as it does not account for R\&D costs, lower production of TPUs compared to NVIDIA GPUs, and the overhead of co-designing TPUs with Broadcom~\cite{reuters2023broadcom}.
The second approach models the equivalent purchase prices of Google TPUs had they been offered for sale, by comparing them to contemporary hardware with similar specifications. We consider this a high-end estimate, because GPU prices include a markup on the cost of developing the chips. We interpolated hardware costs based on price-performance:

\begin{gather*}
\text{TPU effective cost}=
\text{GPU cost}\times\dfrac{\text{TPU performance}}{\text{GPU performance}}\times \text{date adjustment factor}
\end{gather*}
where the date adjustment factor adjusts costs compared on different dates to make them comparable, based on the trend that GPU performance per unit cost improves at a rate of 0.14 orders of magnitude per year.

For the manufacturing cost approach, we estimated the manufacturing cost for the NVIDIA DGX SuperPOD at \$8,665 per GPU. This estimate was informed by \cite{semianalysis2023server} and \cite{kim2023}---calculations are available in `h100\_manufacturing\_cost.ipynb`. After converting this to the TPU cost using the above formula, we divided by the average server-to-chip cost ratio of 1.64 that we estimated from NVIDIA GPU prices (see \Cref{sec:ap2}). The results are listed in \Cref{tab:ta2}. For the equivalent GPU price approach, we found the specifications, release dates, and prices of the most similar non-Google ML GPUs, listed in \Cref{tab:ta3}~\cite{nvidiak80,nvidiap100pcie,nvidiav100sxm2,nvidiaa100sxm4}.

\begin{table}[H]
\centering
\renewcommand{\arraystretch}{1.4}
\resizebox{\textwidth}{!}{%
\begin{tabular}{lccccc}
\toprule
 & \textbf{H100} & \textbf{TPU v1} & \textbf{TPU v2} & \textbf{TPU v3} & \textbf{TPU v4} \\
\midrule
Release date & 2022-09-21 & 2015-05-20 & 2017 & 2018 & 2021 \\
Performance ratio to H100 & 100\% & 5\% & 9\% & 12\% & 28\% \\
Date adjustment factor & 1.00 & 10.66 & 5.38 & 3.90 & 1.48 \\
Server manufacturing cost (per chip) & \$8,665 & \$4,295 & \$4,244 & \$4,200 & \$3,570 \\\midrule
Chip manufacturing cost (estimate) & \$5,346 & \$2,650 & \$2,619 & \$2,591 & \$2,203 \\
Price of equivalent GPU (estimate) & - & \$11,263 & \$9,752 & \$10,742 & \$12,119 \\
Geometric mean & - & \textbf{\$5,463} & \textbf{\$5,054} & \textbf{\$5,276} & \textbf{\$5,176} \\
\bottomrule
\end{tabular}%
}
\vspace{8pt}
\caption{Cost and performance comparison between Google TPUs and the NVIDIA H100.  Performance ratios to the NVIDIA H100 use the same number format and are without sparsity. Our overall estimate of TPU costs is the geometric mean of estimates for the chip manufacturing cost and the price of an equivalent-performance GPU.}
\label{tab:ta2}
\end{table}

\begin{table}[H]
\centering
\renewcommand{\arraystretch}{1.4}
\resizebox{\textwidth}{!}{%
\begin{tabular}{lcccc}
\toprule
 & \textbf{K80} & \textbf{P100 PCIe 16GB} & \textbf{V100 SXM2 32GB} & \textbf{A100 SXM4 40GB} \\
\midrule
\textbf{Release date} & 2014-11-17 & 2016-06-20 & 2018-03-27 & 2020-05-14 \\
\textbf{Performance in TFLOPS} & 8 (FP32) & 19 (FP16) & 125 (FP16) & 312 (FP16) \\
\textbf{Memory} & 24 GB & 16 GB & 32 GB & 40 GB \\
\textbf{Sale price} & \$5,000 (at release) & \$5,699 (at release) & \$11,458 (2018-05-08) & \$15,000 (2020)  \\
 & \$3,700 (2016) &  &  &\$12,500 (2022)  \\
\bottomrule
\end{tabular}%
}
\vspace{8pt}
\caption{Comparison of GPU specifications. By interpolation between GPUs, and their price-performance data, we estimate performance-equivalent prices for TPU versions.}
\label{tab:ta3}
\end{table}

As explained above, we consider the manufacturing costs to be low estimates and the equivalent GPU prices to be high-end estimates of the full production cost. To aggregate the two approaches into a final estimate, we took the geometric mean, as shown in \Cref{tab:ta2}. Each TPU version has an estimated cost (for Google) of about \$5,000.

\subsection{Amortization model}
\label{sec:ap4}

As explained in section \ref{sec:amort}, we estimated the value of the training hardware at the beginning of training as:
\begin{gather*}
\text{Start value per chip} =
\dfrac{\text{Acquisition cost per chip}}{\exp\Big(\big[
\text{Training start date} - \text{Hardware availability date}\big]\cdot r\ln{10}\Big)}
\end{gather*}
where $r$ is a depreciation rate in orders of magnitude per year, and the difference in dates is in years. The hardware availability date depended on the type of hardware. If the hardware was a Google TPU, we used the hardware announcement date. For GPUs, we used a 90-day buffer between the GPU first going on the market and the GPU actually being shipped to the buyer. Our results are robust to variations in this buffer time---see \Cref{sec:depsensapp}.

For the training start date, there were a few known cases---for example, GPT-4 finished training in August 2022~\cite{openai2024gpt4}. Otherwise, we subtracted the training time from the publication date, and then subtracted a further 60 days to account for time spent evaluating the model and writing the paper. Again, our results are robust to variations in this buffer. If the training time was unknown, we used the median of known values in our dataset, which was approximately 33 days. 

The precise way to amortize the training cost through exponential depreciation is:
\begin{gather*}
\text{Amortized training cost} =\text{Start value per chip}\times \text{Number of chips} \times \text{Depreciation during training}
\\
= \text{Start value per chip}\times\text{Number of chips}\times
\Big(1-\exp\big[-\text{Training time} \times r\ln{10}\big]\Big)
\end{gather*}

where training time is in years. However, we could estimate chip-hours more often and more reliably than the training time or the number of chips separately. This is because chip-hours can also be estimated from training compute in FLOP divided by the FLOP/s achieved during training. We used a linear approximation to take advantage of these chip-hour estimates:
\begin{gather*}
\text{Amortized training cost} = \text{Start value per chip}\times\frac{\text{Training chip-hours}}{(365\times 24) \text{ hours/year}}\times r\ln{10}
\end{gather*}

This approximation is valid if $(\text{Training time})\times r\ln{10}$ is small, and this is the case for the training times in our data and our choice of $r=0.14$. In an extreme case, a training time of 1 year results in $1\times 0.14\ln(10) \sim=32\%$ deprecation compared to $1-\exp(-1\times 0.14\ln(10))\sim=28\%$ depreciation. This is not a large difference relative to other sources of uncertainty.

Due to NVIDIA covering defects and component failures under warranty, we concluded that hardware failures are not a significant source of depreciation relative to hardware progress. As one data point, an average of 1 to 2 failures per week occurred when training the BLOOM model on a cluster of 384 NVIDIA A100 GPUs~\cite{le2023bloom}. Even if these were all catastrophic failures, the expected hardware lifetime would be 3.7 years. We expect that NVIDIA replaces or repairs defective GPUs on a faster timescale, which makes the cost of failure small compared to hardware price depreciation.

\subsection{Energy cost estimation}
\label{sec:ap5}

To model the cost of energy consumed by hardware during a training run, we started with the thermal design power (TDP) of the GPU or TPU used for training. We then scaled this up to estimate the TDP of one server. For TPUs, the server scale-up was based on data from Table 1 of \cite{jouppi2021}. For NVIDIA GPUs, we used specifications such as \cite{nvidiadgxh100}.

Next, we converted TDP to the average power actually consumed during training. For TPUs we used an average value of 43\% using data on TDP and average power in~\cite[Table 4]{patterson2021carbon}, as well as data on TDP in~\cite[Table 1]{jouppi2021}. For GPUs we also aggregated multiple sources (\cite{nvidia2023superpod},\cite{patterson2021carbon}, and \cite[p.~133]{barroso2018datacenter}), arriving at an all-things-considered estimate of 75\%.

To account for power consumed by data center power distribution and cooling, we multiplied average server power by the power usage effectiveness. We used a PUE of 1.1 for data centers owned by hyperscalers such as Alphabet and Microsoft, based on~\cite[Table 4]{patterson2021carbon} and a statement by Meta~\cite{meta2024sustainability}. Otherwise, we used 1.25~\cite{semianalysis2024energy}. 

To get the total energy cost of the training run, we multiplied the energy consumption by the average industrial electricity price in the model publication year \cite{electricityprices}.

\subsection{Cloud price selection}
\label{sec:ap6}

To estimate training costs from cloud rental prices, we matched the hardware type and publication date of each ML model with a price from the hardware price database. To do this, we first filtered the database for prices which matched the hardware type and the most likely cloud provider that would be used, with the latter based on the developer of the ML model, e.g. using Google Cloud for any Google lab, and using Microsoft Azure for OpenAI based on their partnership with Microsoft. We then estimated the date of hardware procurement as the publication date minus the training time, and minus a further 2 months to account for preparation before the training run.\footnote{The choice of 2 months was an educated guess. The matching of models to cloud prices was not very sensitive to this choice because the price data was sparse and stable over time.} If the training time was unavailable, we used the median value of approximately 33 days.

Finally, we searched the price database for the price per chip-hour that was dated nearest to the estimated procurement date. We defaulted to the price for a 3-year rental commitment. Based on a few custom quotes we requested from cloud providers, we found that actual cloud computing prices are negotiable and can be substantially lower than publicly listed prices. We concluded that a 3-year commitment price is the closest on average to what developers would be quoted, even if they make a shorter commitment. 

Prices were not available for every specific combination of hardware, cloud provider, and rental date, so we used several fallbacks to select the most closely applicable cloud rental price, for example using nearest prices in time, using prices for similar hardware models, etc. The full procedure is provided at \url{https://github.com/epoch-research/training-cost-trends/blob/main/prices.py#L210-L294}.

\subsection{Accounting for compute used throughout model development}
\label{sec:devcompute}
It is important to consider compute used throughout model development. The cost of experiments, evaluations, and fine-tuning reflects actual costs for developers to research and possibly deploy a useful ML model. This compute is not only important, but significant in scale: we estimate that the ratio of total compute to final training run compute ranges from 1.2x to 4x, with a median of 2.2x.

One source of evidence on the allocation of compute is the training of smaller model sizes for a given architecture. For example, smaller versions of GPT-3 used 4.5e22 FLOP (based on $compute = 6 \times parameters \times tokens$)~\cite[Table 2.1]{brown2020language}. This shows at least 14\% of compute was spent outside the main training run. Similar reasoning for BLOOM reveals about 63\% of compute was used on smaller models~\cite[Table 5]{le2023bloom}.

Another source of evidence is reports of how compute budgets are allocated. For example, the OPT-175B developers estimated total cost at ``roughly 2x higher'' than the largest training run~\cite{zhang2022opt}. Meanwhile, across Meta's AI infrastructure, one estimate in the literature suggested a 1:2 ratio between experimentation and training, where training includes additional hyper-parameter tuning and retraining~\cite{wu2022sustainable}. 

For GPT-3, the true ratio is almost certainly higher than 1.14x due to failures and other experiments. We believe the Meta, BLOOM and OPT-175B cases are the more central examples as they account better for all experiments. So a factor close to 2x seems like a reasonable median estimate. On the high end, it's plausible that several large-scale experiments are necessary before training—say, 4x. We sampled from the range of plausible values using a log-normal distribution. The distribution was defined by a 90\% CI of 1.2x to 4x, leading to a median of 2.2x.

\subsection{Cost uncertainty analysis}
\label{sec:uncertaintyapp}

Our cost estimation methods have many sources of uncertainty, making it important to measure overall uncertainty in the estimates. To do this, we first made a rough estimate of the relative uncertainty in each input variable, based on empirical data. For example, for the overhead of per-GPU server cost relative to single GPU cost we assigned a 90\% credible interval of 1.3x to 2.1x, which is wider than the range of values in our data and from industry sources.\footnote{The three actual values we calculated ranged from 1.54 (P100) to 1.69 (V100). A pre-existing cost breakdown of a DGX H100 implies a ratio of approximately 1.4 (total cost divided by "8 GPU + 4 NVSwitch Baseboard" cost)~\cite{semianalysis2023server}.}

We then used a simulation to sample from distributions over each input variable. The simulation, along with details of the bounds for each input variable, are available at \url{https://github.com/epoch-research/training-cost-trends/blob/main/uncertainty.ipynb}. The simulation used log-normal distributions for all variables except depreciation rate and utilization rate, which used normal distributions. The sampled variables were combined into a sample of final costs, using the relevant formula. The cost sample was then normalized to have a median value of 1. The 90\% CI of this normalized sample represents the relative uncertainty in cost.

The relative uncertainties in cost are listed in \Cref{tab:ta5}. Hardware acquisition cost involves fewer variables with less uncertainty, so estimates are generally accurate within a factor of two for models trained on GPUs, and within a factor of 4 for models trained on TPUs. Meanwhile, amortized hardware CapEx + energy is generally accurate within a factor of three or four for models trained on GPUs, and a factor of five for models trained on TPUs. The cost estimates are most sensitive to the GPU and TPU unit cost (accurate within factors of 2 and 4 respectively) and the training chip-hours (factor of 3).

\begin{table}[H]
\centering
\renewcommand{\arraystretch}{1.4}
\begin{tabular}{l c}
\toprule
\textbf{Cost quantity} & \textbf{90\% CI} \\
\midrule
Hardware acquisition (GPUs) & 0.5x to 2x \\
Hardware acquisition (TPUs) & 0.2x to 4x \\
Amortized hardware CapEx + energy (GPUs) & 0.3x to 4x \\
Amortized hardware CapEx + energy (TPUs) & 0.2x to 5x \\
\bottomrule
\end{tabular}
\vspace{8pt}
\caption{Estimated relative uncertainty in individual cost estimates, for different methods. TPU estimates have larger uncertainty due to the additional uncertainty in estimating their equivalent costs.}
\label{tab:ta5}
\end{table}

\subsection{Ground truth cost comparison}
\label{sec:ap8}

In order to verify that our results are reasonable, we sought to compare our cost estimates with true costs reported by developers and other sources. However, there are very few models where the developers report both the computing resource usage and the total cost. Training costs and compute resources are independently known for BLOOM-176B and OPT-175B, so we compare our estimates with these.

BLOOM-176B was trained on 1,161,261 A100-hours at a throughput of 150 TFLOP/GPU/s at 48\% model FLOPs utilization and a cost of \$3 million (including experiments)~\cite{le2023bloom}. We estimated a cloud compute cost of \$1.99M or an amortized cost of \$0.8M for BLOOM-176B. The accuracy of this estimate depends on how much of the grant was spent on experiments versus the final training run. According to BLOOM’s model page on Hugging Face, the “Estimated cost of training” is the “Equivalent of \$2--5M in cloud computing (including preliminary experiments)”. Preliminary experiments included training smaller BLOOM models. The final training run for the 176B model used 37.24\% of the energy of the BLOOM project~\cite{luccioni2022estimating}; if the total cost of the project was €3M as in the grant description, this implies that BLOOM-176B had a cost of \$1.2M, which is between our two estimates and aligns more closely with the amortized cost approach (\$900K) than the cloud cost approach (\$2M).

OPT-175B was trained for 793.5 hours, at a cost of \$2500/hour as reported in the training logbook~\cite{zhang2022opt}, for a total cost of \$1.98 million. We estimated a cloud compute cost of \$1.5M for the final training run of OPT-175B, which is off by 25\%, and an amortized hardware and energy cost of \$700K, off by 65\%. OPT’s cluster cost rate per hour was likely greater than what we estimate from the quantity of GPUs, or less than the approximate figure mentioned by the developers in the training log. 

\section{Sensitivity analysis}
\label{sec:ap7}
\subsection{Selection of historic frontier models}
\label{sec:ap1}

\begin{figure}\centering
  \includegraphics[width=0.7\textwidth]{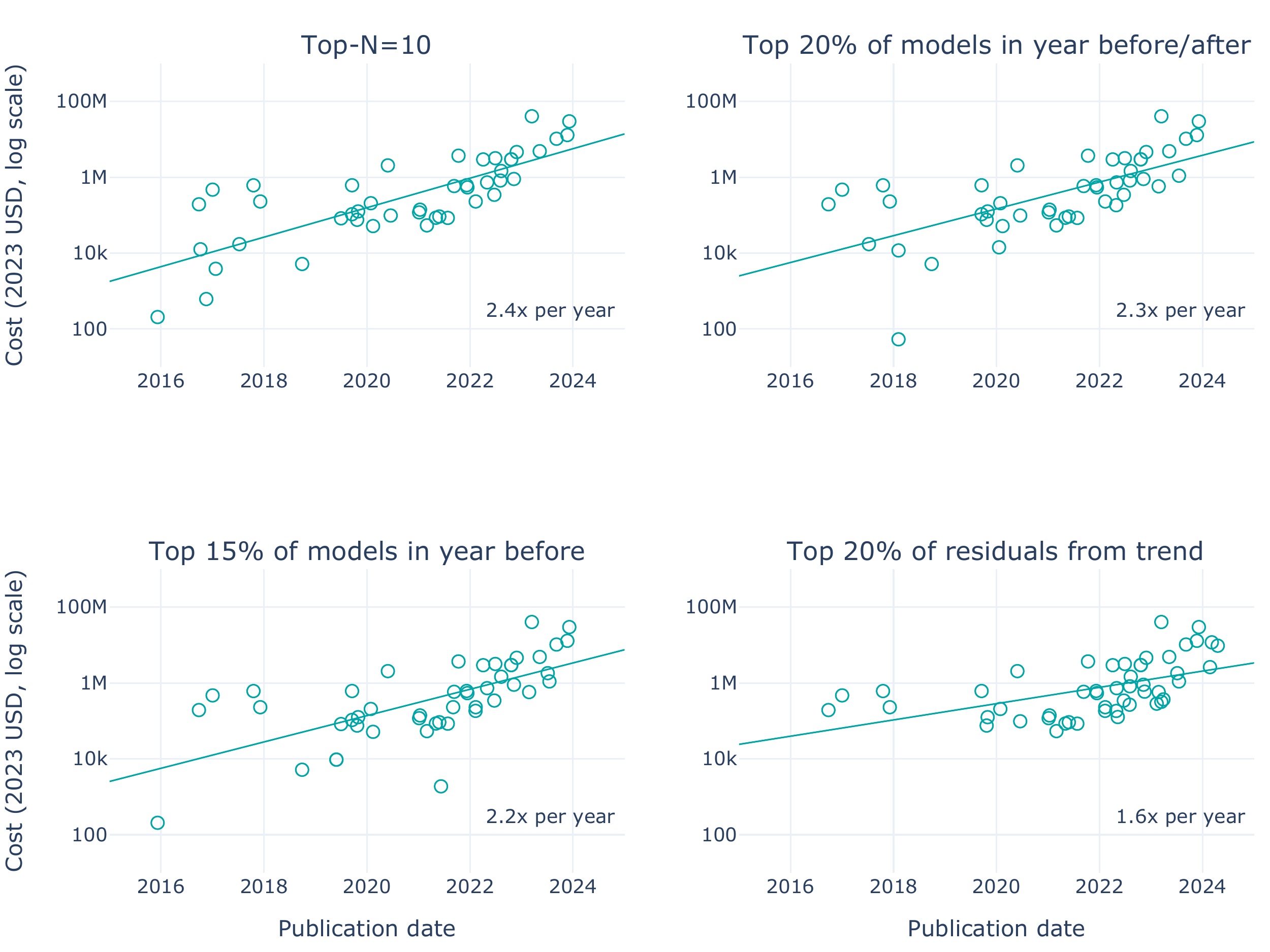}
\caption{Comparison of hardware capex + energy cost regression using different frontier model selection methods. Results are fairly similar across methods, although taking the top 20\% of residuals leads to a flatter trend.}
\label{fig:fa1}
\end{figure}

In order to analyse trends in frontier ML models, we must define what counts as a frontier model at any point in time.\footnote{Models in the database meet one or more of the following criteria: (i) advanced the state of the art on a qualifying benchmark, (ii) at least 1000 citations, (iii) at least one million monthly active users, or (iv) equivalent historical significance~\cite{epoch2023pcdtrends}. However, this means the database includes many models that were far from the frontier of compute.} Our preferred approach is to select models from the database that were in the top 10 most compute-intensive models as of their release date, although we considered others as shown in \Cref{fig:fa1}.

For the most part, different selection approaches gave similar results. The exception was selecting frontier models based on distance from the compute trend. This approach imposes an artificially flat floor on the eligible models. Due to this, it leaves out many earlier models, and produces a flatter cost trend than the other methods.

Our preferred approach has an advantage over alternatives: the selection is more robust to the sampling of our dataset. Approaches based on quantiles, or distance from the historic trend, are influenced by data collected on models \emph{outside} the frontier. Selecting the top-ranked models, in comparison, is merely influenced by whether the dataset contains those frontier models.

\subsection{Varying N in top-N model selection}

\begin{figure}\centering
  \includegraphics[width=0.75\textwidth]{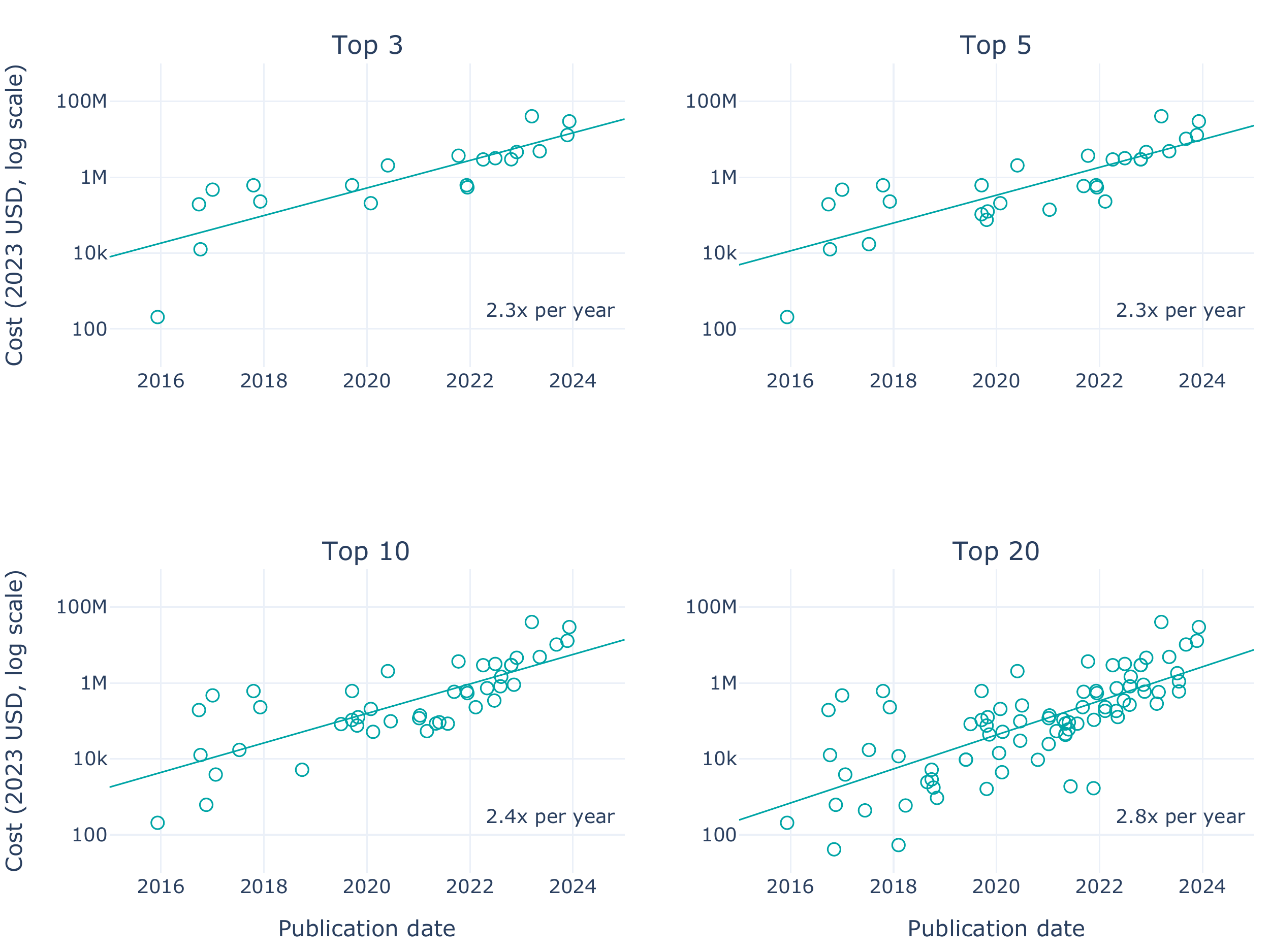}
\caption{Comparison of amortized hardware capex + energy regression for varying top-$N$ selection.}
\label{fig:top-n-comparison}
\end{figure}

When selecting frontier models by the top-$N$ method, there is a question of how to choose $N$. We chose $N=10$ to produce a large enough sample size while still focusing on models near the frontier. The estimated growth rate is moderately robust to the choice of $N$, as it is similar for $N=3$, $N=5$ and $N=20$ (see \Cref{fig:top-n-comparison}).

\subsection{Varying the depreciation of hardware value}
\label{sec:ap73}

The growth in price-performance for ML GPUs running at FP32 precision has been estimated at 0.14 OOMs/year with a 90\% CI of 0.10 to 0.18 OOMs/year~\cite{epoch2023trendsinmachinelearninghardware}. Substituting the lower and upper bounds of that CI for the depreciation rate did not significantly change the growth rate of amortized hardware CapEx + energy. For the lower bound of 0.10 OOMs/year, cost estimates decreased by 15\% on average, while for the upper bound of 0.18 OOMs/year, cost estimates increased by 10\% on average. Note that increasing the depreciation rate has two effects that partially cancel out: 1. the value of hardware at the start of training is smaller, 2. the proportion of value used up by training is larger.

We also tested 0.3 OOMs/year as an extreme case, based on a claim that single-GPU inference performance has improved by $1000\times$ in the last decade~\cite{nvidia2023blog}. This did not significantly change the growth rate either, but it increased costs by an average of 30\%.

\subsection{Varying the time between hardware acquisition and the start of training}
\label{sec:depsensapp}

We tested different estimates of the hardware acquisition date relative to the release date, as well as the training start date relative to the model publication date. These dates affect the time over which hardware value depreciates. To make the depreciation times long, we removed the minimum buffer of 90 days between hardware release and hardware acquisition, and pushed the default training start date back by 15 days relative to the publication date. This decreased the estimated costs by 4\% on average, and did not change the growth rate. Similarly, we tested a short depreciation time by extending the hardware acquisition buffer time to 180 days and bringing the default training start date forward by 60 days. This increased costs by 10\% on average and did not change the growth rate.

\subsection{Excluding equity from R\&D staff costs}
\label{sec:ap74}

\begin{figure}
    \centering
    \begin{subfigure}[t]{1\textwidth}
        \centering
	\includegraphics[width=0.8\textwidth]{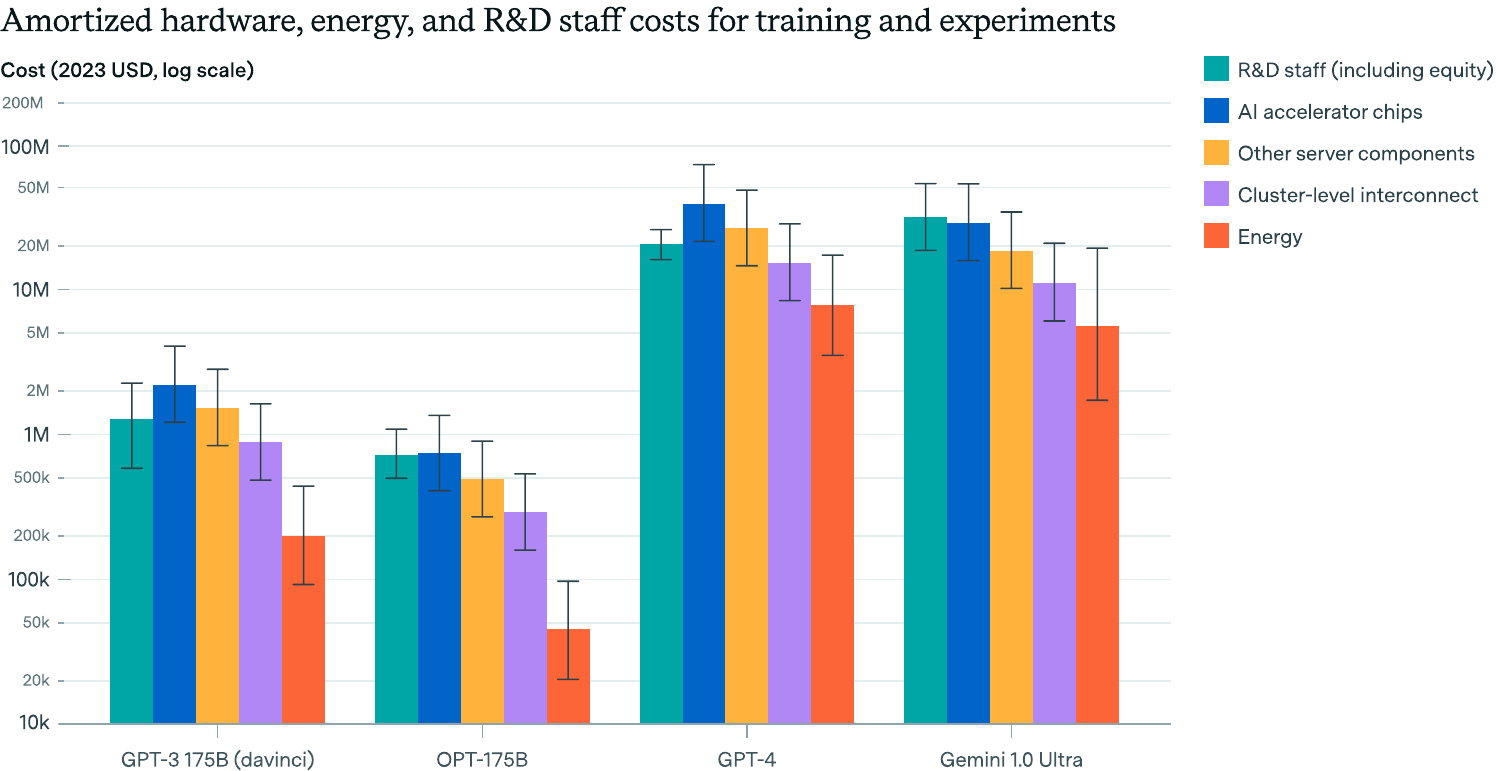}
        \caption{}\label{fig:f11a}
    \end{subfigure}\medskip%

    \begin{subfigure}[t]{1\textwidth}
        \centering
	\includegraphics[width=0.7\textwidth]{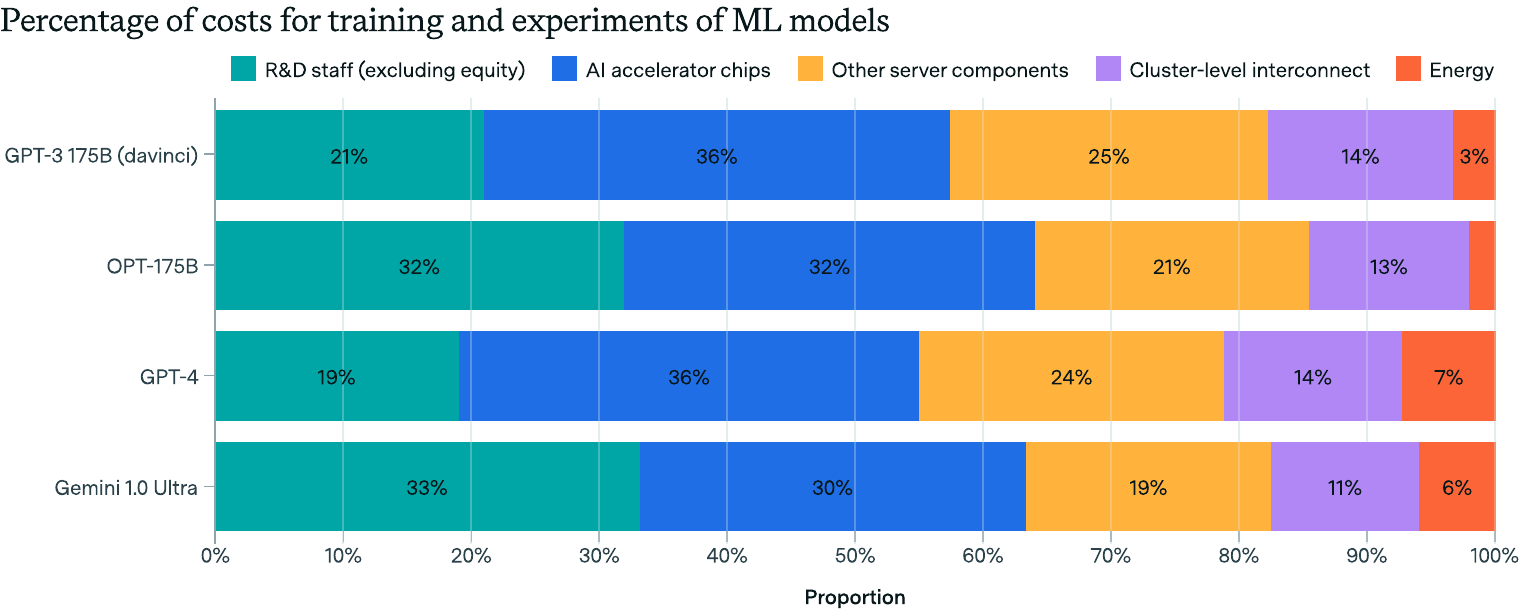}
        \caption{}
        \label{figf11b}
    \end{subfigure}
\caption{(a) Breakdown of total amortized model development costs for selected models, with equity excluded from the R\&D staff cost. Hardware costs are amortized to the total number of chip-hours spent on experiments and training, while R\&D staff costs cover the duration of development from initial experiments to publication. Error bars indicate 90\% credible intervals, while the main bar values are medians. (b) Costs components as a percentage of the total, based on median estimates.}
\label{fig:f11}
\end{figure}

To measure the impact of equity on the total amortized model development cost, \Cref{fig:f11a} shows the cost breakdown with equity \emph{excluded} from the R\&D staff cost. The proportion of cost on R\&D staff decreases from 29--49\% with equity included, to 19--33\% with equity excluded.

\section{Power capacity for model training}
\label{sec:powerapp}

\begin{figure}[H]
\centering
  \includegraphics[width=0.7\textwidth]{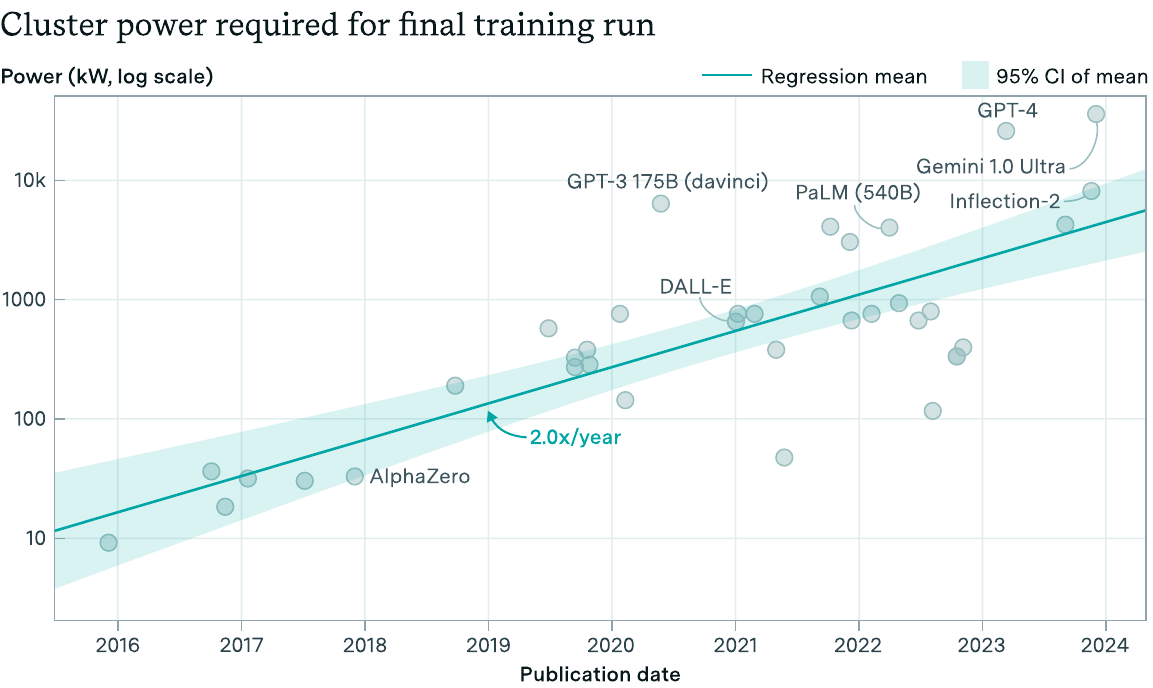}
\caption{The trend in AI compute cluster power (in kilowatts) required to train frontier models over time. Power is calculated as the product of the number of servers, server TDP, and power usage effectiveness.}
\label{fig:fa12}
\end{figure}

\Cref{fig:fa12} shows the trend in the power capacity of the compute cluster needed for training frontier models. This was based on the following formula:
\begin{gather*}
\text{Power capacity (kW)} = \text{Hardware quantity}\times \text{Hardware TDP (kW)}\times\text{Data center PUE}
\end{gather*}
where ``Hardware TDP'' includes all server hardware. We find a growth rate of 2.2x per year (90\% CI: 1.9x to 2.6x). Gemini Ultra has the largest estimated power capacity, at around 35 MW. Projecting the trend forward from Gemini Ultra, the most power-intensive training run would draw 1 GW at some point in 2028. To put this in context, the top ten largest power plants in the United States have a capacity ranging from 3 GW to 7 GW \cite{explained2022electricity}.

\end{document}